\documentclass[twocolumn,showpacs,amsmath,amssymb,pra,floatfix]{revtex4}

\usepackage{bm} 
\usepackage{epsf} 
\usepackage[english]{babel} 
\usepackage{graphicx} 
\def\au{{\rm \,a.\,u.}} 
\def\pb{$\bar{p}$ } 
\def\hmol{H$_2$}%
\def\h2p{H$_{2}^{+}$}%

\def\opp#1{\hat{#1}}

\newcommand{\diff}[1]{\textrm{d}#1} 
\newcommand{\der}[2]{\frac{\textrm{d}#1}{\textrm{d}#2}}

\newcommand{\mean}[1]{\left\langle\,#1\,\right\rangle}

\begin{document}

\bibliographystyle{apsrev} 

%
%
 
\title{Antiproton collisions with molecular hydrogen}   
 
\author{Armin L\"uhr} 
\author{Alejandro Saenz} 
 
\affiliation 
{Institut f\"ur Physik,  
AG Moderne Optik, Humboldt-Universit\"at zu Berlin, Hausvogteiplatz 5-7, 
D-10117 Berlin, Germany.}  

\date{\today} 
 
             
\begin{abstract}\label{txt:abstract} 
Theoretical antiproton and proton cross sections for 
ionization and excitation of hydrogen molecules as well as energy spectra of 
the ionized electrons were  
calculated in the impact-energy range from 8 to 4000\,keV. The cross sections 
were computed with the close-coupling 
formulation of the semi-classical impact-parameter method. The target was 
described using a one-active electron model centered on the target and
assuming a fixed internuclear distance during the collision process. The
dependence of the ionization cross 
sections on the internuclear distance is examined. The present cross sections
are compared with  experimental and theoretical data from the literature. For
impact energies $E  \ge 90$\,keV the obtained results for ionization by
antiproton impact are comparable to the  experimental data while they disagree
for energies $E < 80$\,keV.   
\end{abstract}

\maketitle


\section{Introduction} 
%
%

%
%
%
%
The ionization of atoms and molecules by antiproton ($\bar{p}$) impacts has 
become the subject of great theoretical and experimental interest. The 
motivation for this is twofold. First, there is a fundamental interest in 
collisions involving exotic particles and second, in comparison to proton ($p$)
collisions the sign of the projectile charge is exchanged which opens up the 
possibility to explore interesting physical effects. Further attention is 
drawn to this topic due to the upcoming Facility for Antiproton and Ion
Research (FAIR) \cite{anti:fair}. The international collaborations on atomic
and molecular physics FLAIR \cite{anti:flai} and SPARC \cite{anti:spar}, both
being a part of the FAIR project, intend to investigate antiproton driven
ionization processes and even kinematically complete antiproton collision
experiments \cite{anti:wels07}. These experimental efforts complement the
recent intense activity in experimental \cite{anti:gabr02,anti:amor02} and
theoretical
\cite{anti:zyge01,anti:jons01,anti:jons04,anti:armo05,anti:shar06,anti:voro08}
anti-hydrogen studies.      
 
In the last 15 years a large number of elaborate calculations have been 
carried out for ionization in the simplest one-electron and two-electron 
systems $\bar{p}$ + H 
\cite{anti:well96,anti:schi96,anti:hall96,anti:igar00a,anti:saki00a,anti:pons00,anti:pons01,anti:tong01,anti:tosh01,anti:azum01,anti:saho04,anti:saki04}  
and $\bar{p}$ + He  
\cite{anti:schu03,anti:wehr96,anti:read97,anti:bent98,anti:lee00,anti:kirc02,anti:tong02b,anti:keim03,anti:igar04,anti:saho05,anti:fost08} 
, respectively, and have offered the single-ionization cross sections in 
agreement with experimental results   
\cite{anti:ande86,anti:ande87,anti:ande90,anti:knud92,anti:hvel94,anti:knud95} 
for incident energies $E\ge 50$\,keV. An interesting difference between the 
proton and antiproton impacts was experimentally recognized for the 
double-ionization cross sections of the He atom 
\cite{anti:ande86,anti:ande87,anti:hvel94}, and its origin could be explained 
by theoretical studies \cite{anti:ford94,anti:knud92}. At low energies ($E < 
50$\,keV), however, large discrepancies still remained between the 
experiment and theoretical results both for the single- and double-ionization 
processes. Very recent experimental results for $\bar{p}$ + He collisions at 
energies below 30\,keV could partly resolve this discrepancies 
\cite{anti:knud08}. Theoretical work has also been done for other atomic 
targets, e.g., recently for alkali-metal atoms  \cite{anti:luhr08}.    

For molecular targets measurements were further made for ionization in 
$\bar{p}$ + H$_2$ collisions \cite{anti:ande90a,anti:hvel94}. Again, a notable 
difference between proton and antiproton impacts could be seen for the 
ionization cross sections. While detailed  work 
was done for proton impacts \cite{sct:shah82,sct:shah89,sct:rudd83,sct:rudd85,sct:eliz99}, little is 
investigated about the ionization of molecules by antiproton impact. In the 
case of $\bar{p}$ + H$_2$ collisions a calculation was done by Ermolaev  
\cite{anti:ermo93}, who used an atomic hydrogen target with a scaled nuclear
charge $Z_{\rm n} = 1.09\, Z_{\rm proton}$ in order to mimic the H$_2$
ionization potential. A molecular approach was used
by Sakimoto for calculations of the one-electron system $\bar{p}$ + H$_2^+$  
\cite{anti:saki05}.

The aim of the present work is to examine the \pb + H$_2$ collision process in
some detail and to improve the existing theoretical cross sections. Therefore,
the target molecule is described with an improved model potential
in comparison to the simple model used by Ermolaev in \cite{anti:ermo93}. The
present model provides cross sections for different internuclear distances of
the H$_2$ molecule. Furthermore, the description of the continuum is improved
by expanding the wave functions in B-spline functions as well as by
increasing the number of basis states. The paper 
is organized as follows: Sec.\ \ref{sec:method} considers the
description of the hydrogen molecule as well as the computational
approach. Sec.\ \ref{sec:results} reports on the dependence of the
ionization cross sections on the molecular internuclear
distance. Subsequently, the calculated cross sections for proton and
antiproton impact are presented and compared to literature data. Finally, the
energy spectra of the ejected electrons are taken into account. Sec.\
\ref{sec:summary}  concludes on the present findings and discusses the  
applicability of the used model.  Atomic units are used unless it is
otherwise stated.

%
\section{Method} 
\label{sec:method} 
\subsection{Model potential}
\begin{table}[b] 
  \centering 
  \begin{tabular}{@{\hspace{0.2cm}}l@{\hspace{0.5cm}}l@{\hspace{0.5cm}}c@{\hspace{0.2cm}}}    
%
%
    \hline 
    \hline 
\,\,$R_n$ &\multicolumn{1}{c}{$\alpha$} &  $I_p$[H$_2$]  \\ 
    \hline 
    \,1.0   &0.26255   &0.672753  \\  
    \,1.2   &0.1960    &0.635961  \\  
    \,1.3   &0.1685    &0.619606  \\  
    \,1.4   &0.1440    &0.604492  \\  
    \,1.4487&0.13308   &0.597555  \\  
    \,1.5   &0.1219    &0.590531  \\  
    \,1.55  &0.1127    &0.583959  \\  
    \,1.68  &0.0881    &0.568062  \\  
    \,1.8   &0.0690    &0.554815  \\  
    \,2.0   &0.0434    &0.535499  \\  
    \,2.11  &0.0313    &0.526223  \\  
    \hline 
    \hline 
  \end{tabular} 
  \caption{Values of the model potential parameter $\alpha$ used in this work 
    for different internuclear distances $R_n$ given in \au\ \ For these
    internuclear distances also
    the ionization potential $I_{p}$[H$_2$] for H$_2$ is given in Hartree 
    which is obtained using the H$_2$ ground-state potential-energy curve
    calculated by Wolniewicz  \cite{dia:woln93}.     
    \label{tb:potential-parameter}}  
\end{table} 
%

The target molecule is treated as an effective one-electron system. The
effective electron is exposed to a model potential
\begin{equation} 
  \label{eq:model_potential} 
  V_{\rm mod}(r) = - \frac{Z_n}{r}\, \left(  
                    1 + \frac{\alpha}{|\alpha|}\,\exp\,\left[ 
                                        -\frac{2\,r}{|\alpha|^{1/2}} 
                                      \right ] 
                  \right)\,,  
\end{equation} 
which contains one dimensionless parameter $\alpha$ and the nuclear charge is
$Z_n=+1$. It was proposed by Vanne and Saenz in \cite{sfm:vann08} and further
discussed in \cite{dia:luhr08}. They successfully used
$V_{\rm mod}(r)$ for calculating the ionization and
excitation of H$_2$  molecules in intense ultrashort laser pulses. A
comparison with their results from a full molecular treatment of H$_2$
confirmed the 
applicability of the model for ionization as well as for excitation. 
The model potential reduces to the ionization potential of atomic hydrogen,
$I_p[{\rm H}] = 0.5$\au, for $\alpha \rightarrow 0$. Furthermore, it satisfies
the conditions $V_{\rm mod}(r) \rightarrow -1/r$ for $r \rightarrow \infty$ and
therefore describes, in contrast to the model used by Ermolaev in
\cite{anti:ermo93}, the long-range behavior of the potential correctly.  

The dependence of the ionization potential $I_{\rm mod}(\alpha)$ on the
parameter $\alpha$ for a system described by $V_{\rm mod}$ is determined
numerically in \cite{sfm:vann08} but for $0 \le \alpha \le 0.3$ it can also be
approximated well with the analytic expression 
\begin{equation}
  \label{eq:ion_potential_approx}
  I_{\rm mod}(\alpha) \approx I_p[{\rm H}]+\frac{\alpha}{(1+\sqrt{|\alpha|})}\,.
\end{equation}
 $\alpha$  is chosen such that $I_{\rm mod}$ corresponds to the
ionization potential $I_p[{\rm H}_2]$ of the H$_2$ molecule at a given
internuclear distance $R_n$.  $I_{p}[{\rm H}_2]$ as a function of $R_n$ is
obtained by subtracting the ground-state potential-energy curve of H$_2$ which
was very accurately calculated by Wolniewicz \cite{dia:woln93} from the
ground-state energies of H$_2^+$. The values of $\alpha$ 
used in this work for the various $R_n$ ranging from 1.0\au\ to 2.11\au\ are
given in Table \ref{tb:potential-parameter} together with 
the corresponding ionization potentials. 
 It should be mentioned that within this model the effect 
of anisotropy due to the two nuclei as well as of the second electron is 
solely contained  as a screening of the Coulomb potential. 

A different model potential for the target can be obtained if in Eq.\
(\ref{eq:model_potential}) the parameter $\alpha$ is set to $\alpha
\rightarrow 0$ and the nuclear charge is scaled to $Z_n = Z_{\rm scal} = 1.09$
as it was, for example, proposed by Ermolaev in \cite{anti:ermo93}. In what
follows,  it shall be referred to this special case of $V_{\rm mod}$ as $V_{\rm
  scal}$ which   describes a scaled hydrogen \emph{atom} H$_{\rm scal}$ with
all energy levels shifted according to
\begin{equation}
  \label{eq:energy_scaling}
  \epsilon_{j\,}[{\rm H}_{\rm scal}] = (Z_{\rm scal})^2 \;\epsilon_{j\,}[{\rm
    H}]\,. 
\end{equation}
The ionization potential of H$_{\rm scal}$ is equal to the absolute value of
the ground-state energy
$I_p[{\rm H}_{\rm scal}]=\left|\,\epsilon_{1\,}[{\rm H}_{\rm scal}] \,\right|
= 0.59405$\au\ and corresponds to the ionization potential of a H$_2$ molecule
at an internuclear separation $R_n=1.474$\au\ \   

\subsection{Time propagation}

An approach similar to the one used in this work was already applied
and described in some detail in a previous work \cite{anti:luhr08} dealing
with slow antiproton and proton collisions with alkali-metal atoms. 
The relative motion of the heavy projectiles is approximated by classical 
trajectories with constant velocities ${\bf v}_p$ parallel to the $z$ axis
also known as the impact-parameter method. The 
distance vector $\mathbf{ R}$ between the projectile and the target is given
by 
\begin{equation}
  \label{eq:distance_vector}
  {\mathbf{ R}(t)=\mathbf{b+v}_p}\,t\,, 
\end{equation}
where $\mathbf{b}$ is the impact-parameter vector along the $x$ axis and $t$ 
the time.  The internuclear distance $R_n$ of the two molecular target nuclei
is  held fixed during the collision process.
The time-dependent Schr\"odinger equation  
\begin{equation} 
  \label{eq:tdSE} 
  i \, \frac{\partial}{\partial t} \, \Psi({\bf r},t) =  
  \left (\, \opp{H}_0 + \opp{V}_{\rm int}({\mathbf{ r,R}}(t))\, \right )\, 
  \Psi({\bf r},t)\, 
\end{equation} 
of the target interacting with the projectile is solved. $\opp{H}_0$
is the Hamiltonian of the effective target electron including the model
potential  
(\ref{eq:model_potential})  
\begin{equation} 
  \label{eq:target_hamiltonian} 
  \opp{H}_0 = - \frac{1}{2} \nabla^2 + \opp{V}_{\rm mod}\,. 
\end{equation} 
The time-dependent interaction between the 
projectile with the charge $Z_p$ and the target is given by the time-dependent 
interaction potential  
\begin{equation} 
  \label{eq:interaction_potential} 
  V_{\rm int}({\mathbf{ r,R}}(t)) =  
           \frac{-Z_p}{\left| \mathbf{r - R}(t) \right|} +  
           \frac{Z_p}{|\mathbf{R}(t)|}\,, 
\end{equation} 
where ${\bf r}$ is the spatial coordinate of the effective electron. 
The time-dependent wavefunction  
\begin{equation} 
  \label{eq:total_wavefunction} 
 \Psi({\bf r},t) = \sum_{j}\, c_{j}(t)\, 
                   \phi_{j}(\mathbf{r})\, 
                   \exp[-i\,\epsilon_{j}\,t]
\end{equation} 
is expanded  in eigenstates $\phi_{j}$ of the target Hamiltonian 
$\opp{H}_0$ with the energies $\epsilon_j$. The radial part of $\phi_{j}$ is 
furthermore  expanded in  
$B$-spline functions whereas its angular part is expanded in spherical 
harmonics.   
Substitution of the wavefunction $\Psi$ in Eq.\ (\ref{eq:tdSE}) by its 
expansion given in Eq.\ (\ref{eq:total_wavefunction}) results for a
trajectory with fixed $b$ (cf.\ Eq.\ (\ref{eq:distance_vector})) in a system
of coupled differential equations 
\begin{equation} 
  \label{eq:coupled_equaitons} 
  i\der{}{t}\,c_{j}(t,b) =  
           \sum_{k}  c_{k}(t,b) \, 
           \langle\phi_{j}| \opp{V}_{\rm int}|\phi_{k}\rangle 
           \exp[i(\epsilon_{j} - \epsilon_{k})t]\,.
\end{equation} 
The differential equations 
(\ref{eq:coupled_equaitons}) are solved with the initial conditions 
${c_j({t=-\infty};b)}= \delta_{j\,1}$, where $c_1$ is the coefficient of the 
ground state of the target Hamiltonian $\opp{H}_0$.  
The transition probability $P_{j}(b)$ for a transition into the target state $j$
after a collision with the impact parameter $b$ is given by
\begin{equation} 
  \label{eq:transition_probability} 
  P_{j}(b) = |c_{j}(t=+\infty,b)|^2 
\end{equation} 
and accordingly the probabilities for ionization $ P_{\rm ion}(b)$ and
excitation $ P_{\rm ex}(b)$ by
\begin{eqnarray} 
  \label{eq:ionization_probability} 
  P_{\rm ion}(b) &=& \quad \sum_{\epsilon_j>0} |c_{j}(t=+\infty,b)|^2 \,, \\
  \label{eq:excitation_probability} 
  P_{\rm ex}(b) &=& \sum_{\epsilon_1 < \epsilon_j < 0} |c_{j}(t=+\infty,b)|^2\,.
\end{eqnarray} 

The cross section for the transition into the target state $j$ can be
expressed as
\begin{equation} 
  \label{eq:state_cs} 
  \sigma_{j} =  2\, \pi\, \int  P_{j}(b)\,b\; \diff{b}\,. 
\end{equation} 
For an H$_2$ target the cross section for ionization is defined as 
\begin{equation} 
  \label{eq:ionization_cs} 
  \sigma_{\rm ion} = 2\, \pi\, \int 2 \,  P_{\rm ion}(b) \,b\; \diff{b}
\end{equation} 
and for excitation as
\begin{equation} 
  \label{eq:excitation_cs} 
  \sigma_{\rm ex} =  2\, \pi\, \int 2 \,  P_{\rm ex}(b)\,b\; \diff{b}\,. 
\end{equation} 
The electron-energy spectrum  
\begin{equation} 
  \label{eq:electron_spectrum} 
  S(\epsilon) = \frac{\diff{\sigma_{\rm ion}(\epsilon) }}{\diff{\epsilon}}  = 
           2 \,\sum_{\epsilon_j = \epsilon}  \rho\;\!(\epsilon_j)\; \sigma_j\,, 
\end{equation} 
that is 
the differential cross section for ejecting an electron with an energy
$\epsilon$, can be obtained by using the density of the 
continuum states $\rho$. The factor two in Eqs.\
(\ref{eq:ionization_cs}--\ref{eq:electron_spectrum}) accounts for the two
indistinguishable electrons of the H$_2$ molecule. The introduction of such a
simple factor 
is believed to work as long as independent interactions of the electrons with
the projectile are dominating. This is the case for high projectile
velocities. However, it is unclear to what extend phenomena including
electron-electron-interaction like double-ionization can be described with an
effective one-electron model and furthermore can be extracted from the
determined transition probabilities (Eqs. (\ref{eq:transition_probability}) to
(\ref{eq:ionization_probability})) and therfore are not considered here.

%
%
\section{Results and discussion} 
\label{sec:results} 
%
%
%
%
%
%
%
%
%
%
%
%

%
%
%
%
\subsection{Dependence of the internuclear distance} 
\begin{figure}[t] 
    \begin{center} 
      \includegraphics[width=0.48\textwidth]{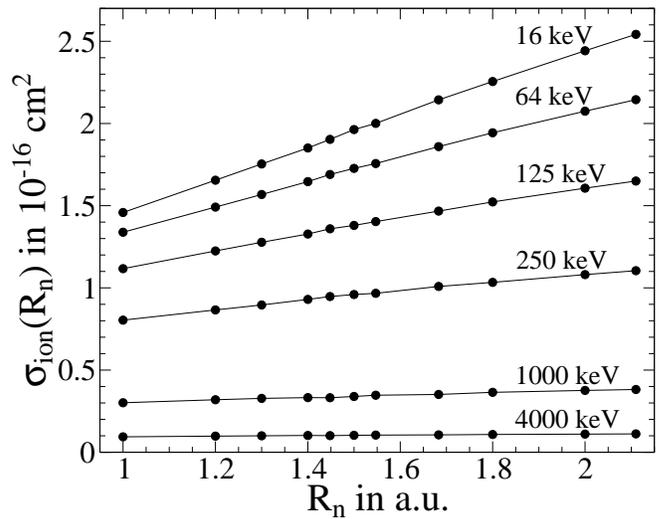} 
      \caption{Ionization cross section $\sigma_{\rm ion}(R_n)$ 
       for $\bar{p}$ + H$_2$ as a function of the internuclear distance $R_n$ at
       $E$\,=\,16, 64, 125, 250, 1000, and 4000\,keV. 
      \label{fig:cs_Rn_H2} } 
    \end{center} 
\end{figure} 

The calculated cross sections depend primarily on the ionization potential
which in turn depends for the H$_2$ molecule on the internuclear distance
$R_n$ (cf.\ Table 
\ref{tb:potential-parameter}). A cross section $\sigma(R_n)$ which depends on
$R_n$ may therefore be defined as 
\begin{equation}
  \label{eq:cs_E_Rn}
  \sigma\:\!(\:\!R_n\:\!) \equiv 
           \sigma\:\!\left(\;\!I_p[{\rm H}_2]\:\!(R_n)\;\!\right)\,.
\end{equation}
%

As has been shown for $\bar{p}$ + H$_2^+$ collisions by Sakimoto 
\cite{anti:saki05} who used a full description of the molecular target ion the
dependence of $\sigma_{\rm ion}(R_n)$ on $R_n$ in the  
range $1.5{\rm \au} \le R_n \le 3.0$\au\ differs for parallel and 
perpendicular orientations. For an orientation of the molecular ion
perpendicular to the collision plane the dependence on $R_n$ is rather
weak. Whereas for an orientation parallel to the trajectory of the projectile
the ionization cross sections increase approximately linearly for larger $R_n$
by more than a factor two in the given $R_n$ range. A stronger orientational
dependence is expected for H$_2^+$ cations compared to a H$_2$ molecule because
of their larger equilibrium internuclear distance. Necessarily, no
molecular-orientation dependence is taken into account in the present
study. Since the present cross sections depend according to  Eq.\
\ref{eq:cs_E_Rn} on the ionization potential which is orientation-independent
the present results are interpreted as orientation-averaged. 
 
Fig.\ \ref{fig:cs_Rn_H2} shows the present $\bar{p}$ + H$_2$ ionization cross
sections $\sigma_{\rm ion}(R_n)$ for the impact energies $E$ = 16, 64, 125,
250, 1000, and 4000\,keV as a function of $R_n$. Ionization cross sections
were calculated for eleven different internuclear distances in the range
$1.0\au\le R_n \le 2.11$\au\ \ It can be seen that for all
impact energies shown here the ionization cross sections increase with larger
$R_n$. This  
can be explained with the decrease of the ionization potential for an
increasing  internuclear distance also shown in Table
\ref{tb:potential-parameter}. The  
dependence of $\sigma_{\rm ion}$ on $R_n$, however, diminishes with higher
impact energies. For energies $E \ge 1000$\,keV $\sigma_{\rm ion}(R_n)$
depends only weakly on  
$R_n$ and increases by about a factor 1.2 in the whole $R_n$ range. For
smaller energies $E 
\le 125$\,keV the dependence on the internuclear distance is much
stronger and for $E=16$\,keV $\sigma_{\rm ion}(R_n)$ increases by more than a
factor 1.7 in the considered $R_n$ range. Furthermore, for all considered
impact energies the cross sections show roughly a linear dependence on
$R_n$. Therefore, one may assume that for all impact energies $E$ considered
in the present investigation the simple relation   
\begin{equation}
  \label{eq:cs_Rn}
  \sigma_{\rm ion}(R_n) =   \sigma_{\rm ion}(\,\overline{R_n}\,) 
                          + \left(R_n - \overline{R_n}\,\right)
                            \left. \der{\sigma_{\rm ion}(R_n)}{R_n}
                            \right|_{\,\overline{R_n}}  
\end{equation}
holds approximately in the examined interval where $\overline{R_n}$ is a fixed
internuclear distance within this interval.

\begin{figure}[t] 
    \begin{center} 
      \includegraphics[width=0.48\textwidth]{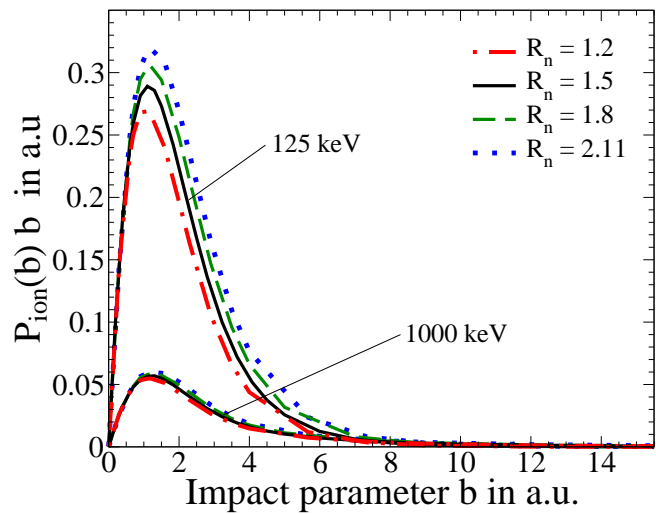} 
      \caption{(Color online) Ionization probability $P_{\rm ion}(b)$ weighted
        with the impact parameter $b$ as a function of $b$ at two
        different antiproton impact energies $E$\,=\,125\,keV and
        1000\,keV. Curves for 
        four different internuclear distances $R_n$ are compared [in \au]: 1.2
        (dot-dashed curve), 1.5 (solid curve), 1.8 (dashed curve), and
        2.11 (dotted curve). 
      \label{fig:Pb_b_Rn_H2} } 
    \end{center} 
\end{figure}

In Fig.\ \ref{fig:Pb_b_Rn_H2} ionization probabilities $P_{\rm ion}(b)$
weighted with the impact parameter $b$ as a function of $b$ are compared for
four different internuclear distances $R_n$\,=\,1.2, 1.5, 1.8, and 2.11\au\ at
two antiproton impact energies $E$\,=\,125\,keV and 1000\,keV. It can be seen
that the curves for the higher impact energy $E$\,=\,1000\,keV differ much
less than those for $E$\,=\,125\,keV in accordance with Fig.\
\ref{fig:cs_Rn_H2}. All maxima of the curves for $E$\,=\,1000\,keV lie
around $b = 1.3$\au\ \ The maxima for $E$\,=\,125\,keV slightly move from $b
= 1.0$\au\ for 
$R_n=1.2$\au\ towards $b = 1.3$\au\ for $R_n=2.11$\au\ and thereby also
increase in height. Whereas, the qualitative behavior of the curves for a
considered impact energy does not change for varying $R_n$.  

In order to determine results which include the rovibrational motion of the
H$_2$ molecule one may use the fact that the cross sections to a given
electronic state can be correctly given using closure (cf.\ \cite{nu:saen97b})
\begin{equation}
  \label{eq:closure}
  \sigma = \int_0^{\infty}
             \sigma(R_n)\,\left|\,\frac{\xi_0(R_n)}{R_n}\,\right|^{\,2} 
             (R_n)^{\, 2} \ \diff{R_n}\,, 
\end{equation}
where $\xi_0(R_n)/R_n$ is the radial nuclear wave function of an H$_2$
molecule in its rovibronic ground state. Clearly, the
integration in Eq.\ (\ref{eq:closure}) leads to a loss of the electron-energy
resolution. The energy information, however, is not relevant for 
integrated cross sections $\sigma$ but for differential cross sections like the
electron-energy spectrum ${\rm d}{\sigma}\,/\,{\rm d}{\epsilon}$.

It is always possible to express
$\sigma(R_n)$ in terms of an (infinite) polynomial in $R_n$ and therefore to
reformulate Eq.\ (\ref{eq:closure}) as 
\begin{eqnarray}
  \label{eq:closure_polynomial_I}
  \sigma &=& \int_0^{\infty} 
                \left( \sum_{k=0}^{\infty} a_k\,(R_n)^{\, k} \right)
                \left|\,\xi_0(R_n)\,\right|^{\,2}  \ \diff{R_n}  \\
 \label{eq:closure_polynomial_II}
         &=& \sum_{k=0}^{\infty} a_k\,\int_0^{\infty}  (R_n)^{\, k} \,
                \left|\,\xi_0(R_n)\,\right|^{\,2}  \ \diff{R_n}  \\
 \label{eq:closure_polynomial_III}
         &=& \sum_{k=1}^{\infty} a_k\,  \mean{(R_n)^{\, k}}  + a_0 \,,
\end{eqnarray}
where $\mean{(R_n)^{\, k}}$ denotes the expectation value of ${(R_n)^{\, k}}$
for the rovibrational ground state of H$_2$. If the cross section
$\sigma(R_n)$ depends on $R_n$ linearly, which is here at least to a good
extend the case, one finds, using Eqs. (\ref{eq:cs_Rn}) and
(\ref{eq:closure_polynomial_III}), the special relation
\begin{equation}
  \label{eq:closure_linear_approximation}
  \sigma = a_1\,  \mean{R_n} + a_0 =   \sigma\left(\mean{R_n}\right)\,,
\end{equation}
i.e., it is sufficient to evaluate the cross section at the expectation value
of the internuclear distance $\mean{R_n}$ of the H$_2$ molecule. The value
$\mean{R_n}=1.448$\au\ has been reported by Kolos and Wolniewicz
\cite{dia:kolo64} and it was used in the present calculations to determine the
ionization and excitation cross sections. 

It may be mentioned that although the 
vibration and rotation of the H$_2$ molecule is taken into 
account a distortion of the molecular vibration and rotation during the
collision with the projectile may possibly lead to a substantial change in
magnitude of the cross 
section. The effect of such a distortion (which is not accounted for in the
present work) on $\sigma$ may be largest for small impact energies where the
cross section depends considerably on $R_n$ as has been shown in
Fig.\ \ref{fig:cs_Rn_H2}. In order to better understand collision processes
involving slow antiprotons ($E<100$\,keV) it would be desirable to fully
include, in an advanced approach, the evolution of the internuclear distance
during the collision. 

%
%
%
\subsection{Ionization of H$_2$ by $p$ impact} 

%
%

 %
\begin{figure}[t] 
    \begin{center} 
      \includegraphics[width=0.48\textwidth]{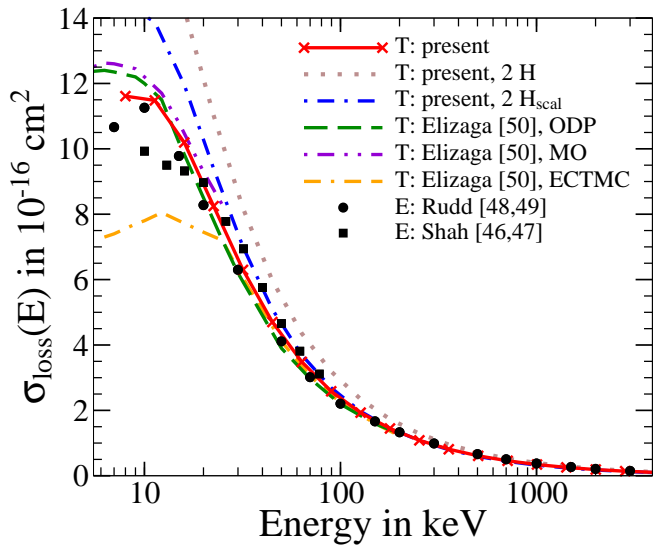} 
      \caption{(Color online)  Electron-loss cross sections $\sigma_{\rm
          loss}$ for $p$ + H$_2$ as a function of the impact energy  $E$. 
       Theory.
       Present results: 
       solid curve, $p$ + H$_2$;  
       dotted curve, $p$ + H multiplied by two; 
       dash--dotted curve, $p$ + H$_{\rm scal}$ multiplied
       by two.
       Elizaga {\it et al.}~\cite{sct:eliz99}:
       dashed curve, optimized dynamical pseudostates method;
       dash--double-dotted curve, molecular orbitals method;
       double-dash--dotted curve, eikonal classical trajectory Monte Carlo
       method.  
       Experimental results: 
       filled circles, Rudd {\it et al.}\ \cite{sct:rudd83,sct:rudd85};
       filled squares, Shah and Gilbody  \cite{sct:shah82,sct:shah89}.
      \label{fig:cs_H2_p} } 
    \end{center} 
\end{figure} 

As has been discussed in a previous work \cite{anti:luhr08} in some detail 
much more effort is needed to bring proton compared to antiproton cross
sections to convergence using the present method. This is in  
particular true for low proton impact energies where electron capture 
becomes the dominant loss
channel of the target electrons. The difficulties in the description of the 
electron capture are mainly due to the use of a one-center expansion of the  
basis around the target which has to be compensated with an enlarged basis
set. The main motivation for the present calculations of proton results is 
given by the need for a comparison of the employed method with an extended
amount of literature since the experimental and theoretical data on antiproton
collisions with 
H$_2$ molecules are still sparse. A one-center expansion around the target,
however, seems to be justified for antiproton collisions in which electron
capture is absent and which are in the focus of this investigation.

The present results for the electron loss of molecular hydrogen in collisions
with protons  
are shown in Fig.\ \ref{fig:cs_H2_p} as a 
solid curve. Also shown are the electron-loss cross sections for \emph{atomic}
hydrogen in a $p$ + H collision multiplied by two. The present data
are compared  with experimental results by Rudd {\it et al.}\
\cite{sct:rudd83,sct:rudd85} and by Shah and Gilbody
\cite{sct:shah82,sct:shah89}.
 
The present findings for H$_{2}$ match the experimental data by Rudd {\it
et al.}\ in the whole energy range. The agreement with the measurements of Shah
and Gilbody is also good except for $E<20$\,keV where their data starts to be
smaller than the results of the present work as well as those of Rudd {\it et
  al.}  
The electron loss cross sections for an \emph{atomic}
hydrogen target in $p$ + H collisions multiplied by two agree well with the
experimental and present data for $E>300$\,keV. With decreasing impact
energies, i.e., with  increasing dependence of the cross sections on the
internuclear distance, the results for $p$ + H get, however, considerably too
large.  


In the theoretical work by Elizaga {\it et al.}~\cite{sct:eliz99}  a similar
model potential 
was used which can be obtained by integrating an effective hydrogen atom-like
charge distribution with Gauss's theorem. This model potential was also
proposed by Hartree in \cite{gen:hart57} for He atoms ($R_n=0$). Cross
sections for the electron loss were calculated for $R_n = 1.4$\au\ \ 
Thereby, the three methods molecular orbitals (MO), optimized
dynamical pseudostates (ODP), and eikonal classical trajectory Monte Carlo
(ECTMC) were used in the calculations and the results are also shown in
Fig.~\ref{fig:cs_H2_p}.    
The cross sections obtained with ODP are very similar to the present
ones. Only for $E < 10$\,keV they are larger than the present data and those
by Rudd {\it et al.}\ The MO approach was applied only at low energies
$E<25$\,keV and leads throughout to similar, though, slightly larger results
than those obtained with ODP. Exactly in the latter energy range the outcome
obtained with ECTMC differs considerably from all other curves whereas for
$E>25$\,keV it matches the experimental and the present results very well.
It can be concluded that the present approach is capable of describing
collisions with \hmol\ targets quite accurately in the considered energy range. 

%
%
%
\subsection{Ionization of H$_2$ by $\bar{p}$ impact} 
\label{sec:results_ionization_pb} 
%
%
%
%
%
%
%
%
%
%
%
   
 %
\begin{figure}[t] 
    \begin{center} 
      \includegraphics[width=0.48\textwidth]{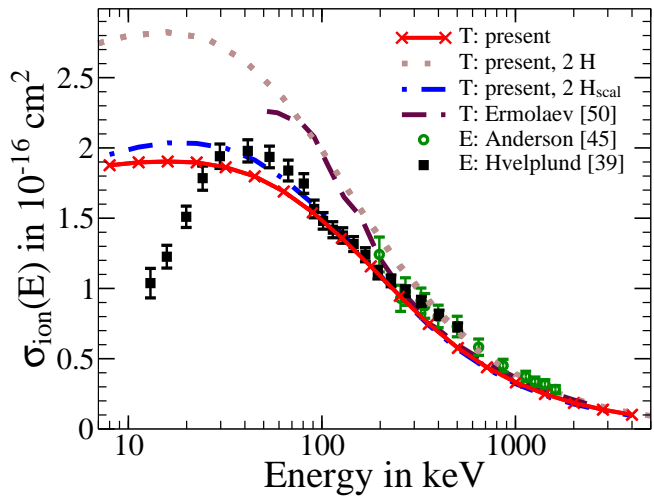} 
      \caption{(Color online)  Ionization cross section $\sigma_{\rm ion}$ 
       for \pb + H$_2$ as a function of the impact energy $E$. Theory:
       solid curve, present results; 
       dotted curve, present
       results for \pb + H multiplied by two; dash-dotted curve, present
       results for \pb + H$_{\rm scal}$ multiplied by two;  dashed curve,
       results for \pb + H$_{\rm scal}$ multiplied by two by 
       Ermolaev  \cite{anti:ermo93}. 
       Experiment: circles, Anderson {\it et al.}\
       \cite{anti:ande90a}; squares, Hvelplund {\it et al.}\
       \cite{anti:hvel94}.  
       \label{fig:cs_H2} } 
    \end{center} 
\end{figure} 

The present results for the ionization of molecular hydrogen by antiproton 
impact 
are shown in 
Fig.\ \ref{fig:cs_H2} as solid curve. Also shown are the ionization cross
sections for antiproton collisions with \emph{atomic}   
hydrogen multiplied by two. The results are compared with calculations by 
Ermolaev \cite{anti:ermo93} and experimental data for non-dissociative
ionization by Anderson {\it et al.} \cite{anti:ande90} as well as data of a
subsequent measurement by  Hvelplund {\it et al.} \cite{anti:hvel94}. As has
been suggested by Hvelplund {\it et al.} in  
\cite{anti:hvel94} the data for impact energies below 200\,keV of their 
earlier measurement \cite{anti:ande90} are omitted in Fig.\ 
\ref{fig:cs_H2}. The data for $E < 200$\,keV of their first experiment are 
generally some $10\%$ larger than those in \cite{anti:hvel94} but have a 
considerably lower accuracy.

For high impact energies $E \ge 1000$\,keV all theoretical curves coincide and
also agree with the experimental data. For lower energies ($400\,{\rm keV} < E
<  1000$\,keV) the ionization cross sections for atomic hydrogen start to
differ  from both theoretical results for a hydrogen molecule. However, at these
energies the atomic results seem to describe better the experimental data. In
the energy  
regime from 250\,keV down to 90\,keV the theoretical cross sections by
Ermolaev approach those of the \pb + H calculation which differ significantly
from the measured cross sections. The experimental data are, however, well 
described by the present \pb + H$_2$ cross section in this energy
regime. Though, the strong variation of the experimental data around 85\,keV
is not  followed by the smooth curve of the present results. While the
magnitude of the present cross sections is comparable to the  
experimental data down to 20\,keV the functional behavior of both, experimental 
and present curve, starts to differ for $E<50$\,keV. Here, the present \pb + 
H$_2$ curve possesses a similar characteristic as two times the cross  
sections of the hydrogen atom  but with smaller magnitude because of the
larger ionization potential of the molecule. The slope of the present cross
sections at these low energies may indicate the lack of two-electron effects
in the target description. The experimental data, on the  other hand, show a
behavior very similar to that of the single ionization of helium also measured
with the same  
experimental set-up by  Hvelplund {\it et al.} \cite{anti:hvel94}. Very 
recently the same authors published another measurement of the single
ionization cross section for \pb + H$_2$ in the energy range $3\,{\rm keV} < E
< 25$\,keV \cite{anti:knud08} which revealed that their helium single
ionization cross sections in \cite{anti:hvel94} are too small for the lowest
measured energies. It may be an interesting question whether the same is true
in the case of the \pb + H$_2$ ionization cross sections as suggested by the
present results. Therefore, it would be worthwhile to initiate a further
attempt to  measure  \pb + H$_2$ cross sections at low antiproton energies.

An effective one-electron description with a fixed internuclear 
distance seem to be sufficient to describe non-dissociative ionization cross
sections for \pb + H$_2$ at high energies. But it is unclear how strong
the influence of two-electron effects and the variation of the internuclear
distance is at intermediate and low energies. Since the energy regime around
and below the maximum of the ionization cross section is considered to contain 
interesting physical effects a full quantum mechanical treatment of the 
target molecule would be desirable. It should be mentioned, however, that such
an approach is very demanding.  

%
%
%
\subsection{Excitation of H$_2$ by $\bar{p}$ impact} 
%
%
%
%
%
%
 %
\begin{figure}[t] 
    \begin{center} 
      \includegraphics[width=0.48\textwidth]{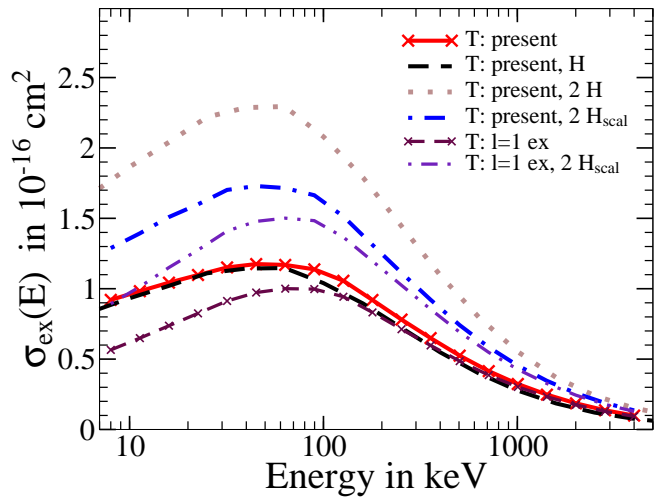} 
      \caption{(Color online)  Excitation cross section $\sigma_{\rm
          ex}$ for \pb + H$_2$ as a function of the impact energy $E$. Theory: 
       solid curve, present results; 
       long-dashed curve, results for \pb + H; dotted curve, 
       same results for \pb + H multiplied by two; dash-dotted curve, results
       for \pb + H$_{\rm scal}$ multiplied by two. Cross sections for
       excitation into $l=1$ states. Theory: 
       thin short-dashed curve, present results; 
       thin dash-double-dotted curve, results for \pb + H$_{\rm scal}$
       multiplied by two.    
       \label{fig:cs_H2_ex} } 
    \end{center} 
\end{figure} 

The present excitation cross sections for \pb + H$_2$ 
are shown in Fig.\ \ref{fig:cs_H2_ex} as solid curve. Also shown are results
for antiproton collisions with \emph{atomic} hydrogen and the same
\emph{atomic} cross sections multiplied by two. To the best of the authors'
knowledge there are no data in literature to compare these results with.  

Due to the experiences with the ionization cross sections one may 
estimate the range of validity of the excitation cross sections presented here 
to be about $100\,{\rm keV} \le E \le 4000$\,keV. Comparing the results for
ionization and excitation in \pb + H$_2$ collisions one can say that
$\sigma_{\rm ex}$ is smaller than $\sigma_{\rm ion}$ for impact energies
$E<1000$\,keV and that both are practically the same for larger energies. The
maximum of $\sigma_{\rm ex}(E)$ lies around $E=58$\,keV and therefore at a
higher energy than the maximum for ionization. 

The excitation cross sections for molecular hydrogen can also be compared with
the results for \emph{atomic} hydrogen. Fig.\ \ref{fig:cs_H2_ex} clearly shows
that the naive assumption an H$_2$ molecule is essentially composed of two
\emph{independent} hydrogen atoms yields excitation cross
sections which are obviously different from those which were 
obtained with the model potential $V_{\rm mod}$ given in Eq.\
(\ref{eq:model_potential}). Only for high impact energies both curves get
close to each other. On the other hand, it is 
interesting to observe that the excitation cross sections for a \emph{single}
hydrogen atom seem to be much more in accordance with the
present molecular $\sigma_{\rm ex}$. Both cross sections show the same
behavior and have practically the same values in the considered energy
range. This similarity for atomic and molecular hydrogen targets was 
evidently not found in the case of ionization in Sec.\
\ref{sec:results_ionization_pb}.

%
%
%
\subsection{Electron spectra} 
\begin{figure}[t] 
      \vspace{-1.5cm}

      \includegraphics[width=0.490\textwidth]{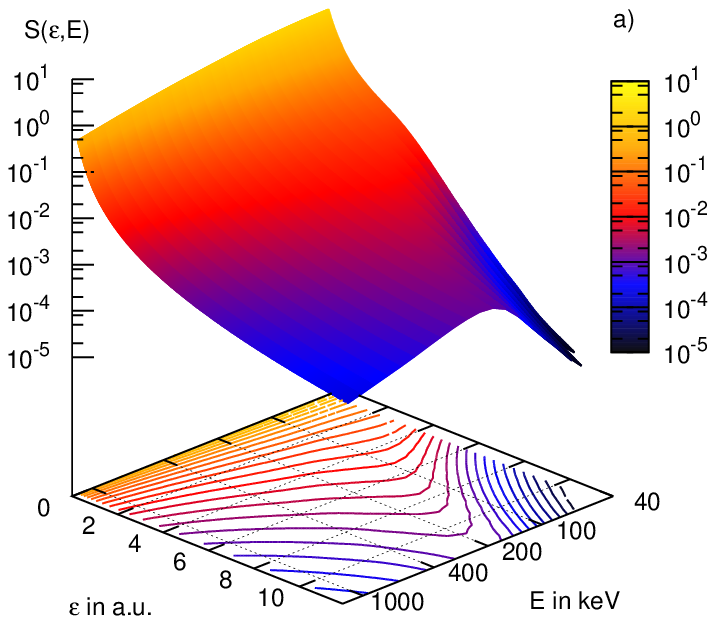}
      \vspace{-2.5cm}

      \includegraphics[width=0.490\textwidth]{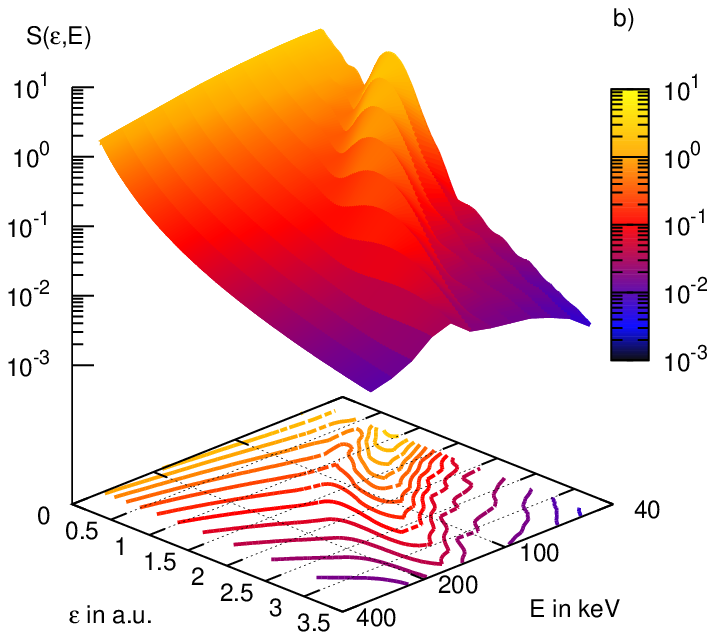}

      \vspace{-1cm}
      \caption{(Color online) Electron-energy spectra surface
        $S(\epsilon,E) = \diff{\sigma}(\epsilon,E)\, /\,\diff{\epsilon}$ given
        in 10$^{-16}$\,cm$^2$\,/\,\au\ 
        as a function of the electron energy $\epsilon$ in hartree and the
        impact energy of the antiproton $E$ in keV. \,a) $\bar{p}$ + H$_2$; \,b)
        $p$ + H$_2$.
      \label{fig:spect_H2_3d} } 
\end{figure} 
\begin{figure}[t] 
    \begin{center} 
      \includegraphics[width=0.48\textwidth]{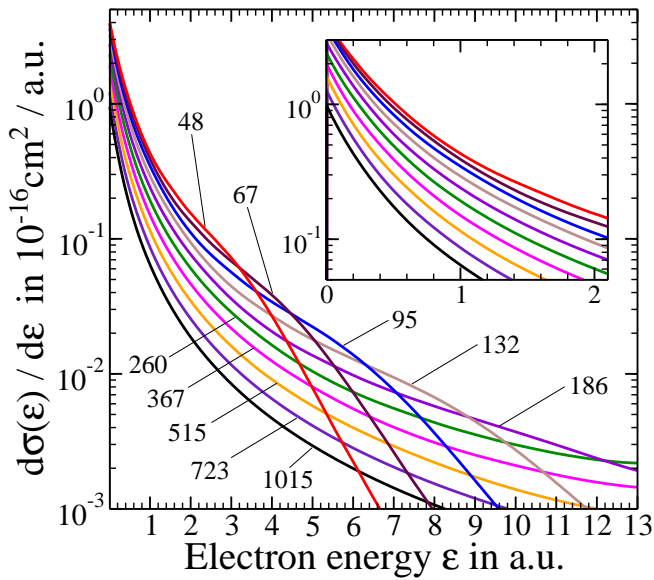} 
      \caption{(Color online) Electron-energy spectra curves
        $S(\epsilon) = \diff{\sigma}(\epsilon)\, /\,\diff{\epsilon}$ for
        $\bar{p}$ + H$_2$ as a function of the electron energy $\epsilon$ at
        $E$\,=\,48, 67, 95, 132, 186, 260, 367, 515, 723, and 1015\,keV. 
        The inset shows the spectra for the  range
        0\,$\le$\,$\epsilon$\,$\le$\,2\au\ without scaling the $y$ axis. The
        curves in the inset are order accordingly to the impact 
        energy $E$. The uppermost curve belongs to the smallest (48\,keV) and
        the lowest  curve to the highest (1015\,keV) impact energy $E$.
      \label{fig:spect_H2} } 
    \end{center} 
\end{figure} 
%
%
%
%
The electron-energy spectra $S(\epsilon,E) = \diff{\sigma}(\epsilon,E)\,
/\,\diff{\epsilon}$ of ionized electrons in a \pb + H$_2$ collision are 
presented in Fig.\ \ref{fig:spect_H2_3d}a as a function of the electron energy
$\epsilon$ and the impact energy of the antiprotons $E$. As has been mentioned
before, the disadvantage of the closure approach in Eq.\ (\ref{eq:closure})
lies in the loss of the 
detailed electron-energy information of the transitions probabilities which is
of relevance to the electron-energy spectra (cf.\ \cite{nu:saen97b}). Therefore,
the presented results may be interpreted as electron spectra for a fixed
internuclear distance rather than including the full rovibrational motion of the
nuclei as it is the case for the integrated cross sections which have been
discussed before. The electron spectra are calculated for a wide 
electron-energy range $0<\epsilon<12\au$\ and for different impact energies of
the antiproton ranging from 48\,keV to 1015\,keV. The contour plot on the
bottom of Fig.\ \ref{fig:spect_H2_3d}a shows 
the corresponding level curves and gives therefore information on the gradient
of the spectra surface. It can be seen that within the whole impact-energy
range the electron spectra decrease smoothly and monotonically for increasing
$\epsilon$. Considering small electron energies $\epsilon<2$\au, the spectra
fall off strongly in view of the logarithmic scale for all impact
energies. Within this $\epsilon$ interval, Fig.\ \ref{fig:spect_H2_3d}a shows
that the smaller the impact energies $E$ the larger the values of
$S(\epsilon,E)$. However, for larger $\epsilon$ this uniform trend
starts to cease. For $\epsilon>4$\au\ the overall decrease becomes
weaker. Though, the electron spectra for small $E$ start to decrease again
very strongly where the fall-off of the spectra is the steepest for the
smallest $E$. 
Consequently, in the intervals of $\epsilon$ and $E$ considered here, the
largest value of $S(\epsilon,E)$ for a given $\epsilon$ moves from $E=48$\,keV
at $\epsilon = 0$ to $E\approx 200$\,keV at $\epsilon=12$\au\

Cuts $S(\epsilon)$ of the same electron-spectra surface for ten different
antiproton impact energies $E$\,=\,48, 67, 95, 132, 186, 260, 367, 515, 723,
and 1015\,keV are presented in Fig.\ \ref{fig:spect_H2}. The inset shows the
same $S(\epsilon)$ curves in an interval of small electron energies
0\,$\le$\,$\epsilon$\,$\le$\,2\au\ Thereby, 
the scaling of the $y$ axis of the inset is kept as it is the main graph. The
ordering of the $S(\epsilon)$ curves in the inset is according to their impact
energy $E$, i.e., the uppermost curve is the one for the smallest (48\,keV)
and the lowest curve the one for the largest (1015\,keV) impact energy $E$. It
can be seen that no crossings of the electron-spectra curves $S(\epsilon)$
occur in this low electron-energy regime.  

In contrast to the behavior for small $\epsilon$ shown in the inset the
$S(\epsilon)$ curves start to cross each other at higher electron energies. 
The curve for $E=48$\,keV starts to fall off much
steeper than the other  $S(\epsilon)$ curves for $\epsilon > 3$\au\ and 
therefore crosses all lower lying curves. Its first crossing takes place at
$\epsilon \approx 3.19$\au\ while its last crossing occurs at $\epsilon
\approx 6.13$\au\  with the curve for $E=1015$\,keV. 
The other electron-energy curves for higher antiproton impact
energies share the same characteristics, namely, that the curve with the
largest values of $S(\epsilon)$ in a certain $\epsilon$ range starts to fall
off steeper than all other lower lying spectra curves for higher impact
energies. Though, with increasing impact energies $E$ the decline of the
$S(\epsilon)$ curves starts at larger $\epsilon$ and gets less steep.

\begin{figure}[t] 
    \begin{center} 
      \includegraphics[width=0.48\textwidth]{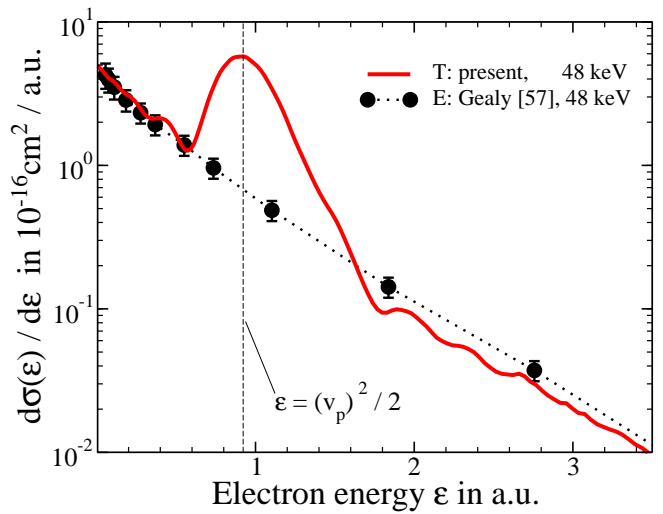} 
      \caption{(Color online) Electron-energy spectra curve
        $S(\epsilon) = \diff{\sigma}(\epsilon)\, /\,\diff{\epsilon}$ for
        $p$ + H$_2$ as a function of the electron energy $\epsilon$ at
        $E$\,=\,48\,keV. 
        Theory: solid curve, present results for electron loss. Experiment:
        filled circles, results for ionization, Gealy {\it et al.}\
        \cite{sct:geal95a}. The energy $\epsilon$ of an electron with the
        velocity of the proton $v_p$ is indicated by the vertical line. 
      \label{fig:spect_H2_p} } 
    \end{center} 
\end{figure}

%
%
%
%
For comparison to the antiproton results in Fig.\ \ref{fig:spect_H2_3d}a an
electron-energy spectra surface $S(\epsilon,E)$ is also presented for $p$ +
H$_2$, i.e., for proton impact, in Fig.\ \ref{fig:spect_H2_3d}b. The electron
spectra are given for the electron-energy range  $0<\epsilon<3.5\au$\ and for
proton impact energies from 48\,keV to 310\,keV. In general the values of
$S(\epsilon,E)$ decrease for larger $\epsilon$. However, the most striking
feature of Fig.\ \ref{fig:spect_H2_3d}b, in contrast to the case of antiproton
impact, is the existence of local maxima of 
the spectrum curves $S(\epsilon)$ for a given impact energy $E$ which
are also visible in the contour plot on the bottom of the figure. 
The position of 
the peaks of $S(\epsilon,E)$ varies with the impact energy $E$. At the center
of the maxima the ratios of the two energies $E$ and $\epsilon$ are such that
the classical velocities of the proton $v_p$ and of the electron $v_e$ are
equal, i.e.,   
\begin{equation}
  \label{eq:spectra_max}
  \sqrt{\frac{2\,E}{M_p}} = v_{p} = v_{e} = \sqrt{{2\,\epsilon}}\,,
\end{equation}
which can be reformulated as   
\begin{equation}
  \label{eq:spectra_max_e_E}
  \epsilon = \frac{E}{M_p} =  \frac{1}{2}\,(v_{p})^2\,,
\end{equation}
where $M_p$ is the proton mass. 
The accuracy of this statement  is demonstrated in Fig.\ \ref{fig:spect_H2_p}
where the present $S(\epsilon)$ spectrum curve for protons with an impact
energy $E=48$\,keV, i.e. $v_p=1.3856$\au, is shown as solid curve. The maximum
of $S(\epsilon)$ is located at $\epsilon=\frac{1}{2}\,(v_{p})^2=0.96$\au, also
indicated by the vertical line. 

The occurring maxima can be explained with the simple picture
of the electron-capture process where the electron is captured by the proton
and moves basically with the momentum of the
projectile. Therefore, the velocity of the captured electron relative to the
H$_2$ molecule is given by the velocity of the projectile, namely the
proton, as well as the electron velocity relative to the moving rest frame of
the projectile. Since both contributions to the electron momentum can be
oriented in different directions the peaks of the electron spectra
$S(\epsilon)$ are centered around the energy $\epsilon$ which corresponds to a
free electron with the velocity of the projectile, cf.\ Eq.\
(\ref{eq:spectra_max_e_E}). It may be mentioned that the capture peaks get much 
less pronounced for higher impact energies. This is, first, due to the
diminishing probability of capture for larger $E$ and, second, due to a
broader $\epsilon$ distribution of the captured electrons.

If the discussed maxima of $S(\epsilon,E)$ in Fig.\ \ref{fig:spect_H2_3d}b are
removed one is left with a smoothly decreasing electron-spectra surface
$S_{\rm nc}(\epsilon,E)$  for 
increasing $\epsilon$ which is similar to the one for antiproton impact
in Fig.\ \ref{fig:spect_H2_3d}a. This modified
$S_{\rm nc}(\epsilon,E)$ for proton impact may be interpreted as the
electron-energy spectrum surface where the electron capture by the 
projectile is excluded. In Fig.\ \ref{fig:spect_H2_p} the present
$S(\epsilon)$ curve for a proton impact energy $E=48$\,keV is compared with
experimental data by Gealy  {\it et al.}\ \cite{sct:geal95a} for which capture
is excluded.  
The comparison shows that except for the $\epsilon$ regime where capture is
the dominant process, i.e. $0.6>\epsilon >1.6$, the present results agree 
with the experimental data though they underestimate the experimental findings
 for high electron energies. The integral of the difference between
the present and the experimental curve over $\epsilon$ ( $\approx 2.2\times
10^{-16}$\,cm$^2$) yields the capture probability for $E=48$\,keV  calculated
by Shingal and Lin \cite{sct:shin89} ($\approx 2\times 10^{-16}$\,cm$^2$) to a
good extend.  The reason for the structures of the
theoretical curve for energies close to the capture peak is not exactly
known. It is likely that they originate from the finiteness of the numerical
description. 


%
%
%
\subsection{Comparison of the models  $V_{\rm mod}$ and $V_{\rm scal}$} 
%
%

%
%
To the best of the authors' knowledge the only existing calculation for \pb +
H$_2$ collisions was performed by Ermolaev who used the potential  $V_{\rm
  scal}$ to describe  the target which is basically an \emph{atomic} hydrogen
target H$_{\rm scal}$ with a scaled nuclear charge $Z_n=1.09$.
His results, shown in Fig.\ \ref{fig:cs_H2}, are not conform with the
experimental data and the present findings. In order to find out why the one
model $V_{\rm mod}$ yields much better results than the other and whether the
same disagreement occurs also for $p$ impact the same cross sections were
calculated again for $p$ and \pb collisions but using $V_{\rm scal}$ in order
to describe the target. 
The resulting cross sections for \pb and $p$ impact multiplied by the factor
two are also shown in Figs.\ \ref{fig:cs_H2_p} to \ref{fig:cs_H2_ex} as
dash-dotted curves. In what follows three remarks shall be made concerning the
results of the calculations using the scaled hydrogen potential $V_{\rm scal}$.
 
First, it is obvious that the present results 
for ionization in \pb + H$_{\rm scal}$ collisions shown in Fig.\
\ref{fig:cs_H2} clearly deviate from those of Ermolaev \cite{anti:ermo93}. It
is astonishing that the latter results by 
Ermolaev better match the present data for \emph{unscaled atomic} hydrogen
than the present and experimental data for molecular hydrogen. No detailed
information is given in 
\cite{anti:ermo93} concerning the employed basis set and the convergence of the
calculations. In very similar studies 
by Ermolaev \cite{sct:ermo90,anti:ermo90}, however, a two-center Slater-type
orbital expansion with 51 basis functions were applied to describe the collision
process between $p$ and \pb projectiles and (unscaled) hydrogen atom
targets. It may be mentioned that the quality of the continuum description in
the calculations by Ermolaev has been put into question by other authors
\cite{anti:hvel94,anti:tosh93}, especially in the so-called 'polarization
region', i.e.\ between $\sim\!70$ and 500\,keV.

Second, the present ionization cross sections  using $V_{\rm mod}$ and $V_{\rm
  scal}$ as target potentials both yield, especially for $E>100$\,keV,
comparable results which can be seen in Figs.\ \ref{fig:cs_H2_p} and
\ref{fig:cs_H2} for $p$ and \pb  impact, respectively. Deviations become
visible for $E<100$\,keV. The similar behavior may be
explained with their comparable ionization potentials $I_{\rm mod} = 0.5976$\au\
at $R_n=\mean{R_n}$ and $I_p [{\rm H_{scal}}]=0.5945$\au\ 

Third, the present cross sections for excitation in Fig.\ \ref{fig:cs_H2_ex}
differ, however, considerably for $V_{\rm mod}$ and $V_{\rm scal}$. To the
best of the authors knowledge no literature data exists to compare the present
results with and therefore to judge which of both models is superior in
describing excitation of H$_2$ molecules. On the 
other hand it has been observed that the excitation cross sections of
alkali-metal atoms depend considerably on the energy difference $\Delta
\epsilon$  between the energetically lowest dipole-allowed $p$ states and the
ground states \cite{anti:luhr08}, i.e., the excitation energy. 
In this context, it shall be noted that also in the present investigation the
dipole-allowed transitions from the ground state to the bound states 
with angular momentum $l=1$, namely the $p$ states, play, especially for
$E>100$\,keV, a dominant role. The cross sections 
for excitations into bound  $l=1$ states are also shown in Fig.\
\ref{fig:cs_H2_ex} for  $V_{\rm mod}$ as thin short-dashed curve and $V_{\rm
  scal}$ as thin dash-double-dotted curve.  
%
%
In \cite{dia:luhr08} the excitation energies $\Delta \epsilon$
for dipole-allowed transitions are compared for the two
models $V_{\rm scal}$ and $V_{\rm mod}$. Therein, it turns out that the
excitation energies calculated with $V_{\rm scal}$ are smaller than those for
$V_{\rm mod}$ 
throughout the $R_n$ range which is considered here. The substantial
differences of the excitation cross sections in Fig.\ \ref{fig:cs_H2_ex} can
therefore be understood by considering the diversity of the $\Delta \epsilon$
curves for both model potentials, namely, the lower excitation energies for
$V_{\rm scal}$ lead to larger excitation cross sections compared to those of
$V_{\rm mod}$.  

In order to find out how well the excitation is described by the employed
models the excitation-energy curves obtained with $V_{\rm mod}$ can also be 
compared with the $\Delta\epsilon$ curves of exact H$_2$ calculations.
For such a comparison it has to be considered that the H$_2$ molecule can be
oriented arbitrarily in a collision. Therefore, transitions from the H$_2$
ground state to states with the molecular symmetries $^1\Sigma_u$ and
$^1\Pi_u$ are both dipole-allowed. The molecules 
are oriented statistically $2/3$ perpendicular and $1/3$ parallel to the
projectile momentum. Consequently, the sum of accordingly weighted
excitation energies, namely,
$2/3\,\Delta\epsilon\,(^1\Pi_u) + 1/3\,\Delta\epsilon\,(^1\Sigma_u) $, should be
compared to the excitation energy of the model $V_{\rm mod}$ for transitions
into $p$ states. The comparison for the whole $R_n$ range considered in the
present work was done in \cite{dia:luhr08} and yielded a good agreement
between the considered excitation energies of the model and the exact H$_2$
molecule.  
Therefore, it is reasonable to assume that also the present excitation cross
sections calculated with the model $V_{\rm mod}$ are superior to those
calculated with $V_{\rm scal}$. For high impact energies the results of the
model $V_{\rm mod}$  may even match the excitation cross sections for exact
H$_2$ molecules completely.  

It shall be emphasized that $V_{\rm mod}$ depends only on one parameter
$\alpha$ which is determined by the ionization potential. There are no
additional parameters in order to fit the energies or wave functions of excited
states. Therefore, it is remarkable that in spite of the simplicity of the model
potential $V_{\rm mod}$ it is possible to reproduce cross sections reasonable
well for ionization and excitation of H$_2$ molecules in strong laser fields
\cite{sfm:vann08} as well as in collisions with antiprotons.

%
%
%
%
%
%
%
%

%
%
%
%
\section{Conclusion} 
\label{sec:summary} 
Time-dependent close-coupling calculations of ionization and excitation cross 
sections for antiproton and proton collisions with molecular hydrogen have 
been performed in a wide impact-energy range from 8 to 4000\,keV.  
The target molecule is treated as an effective one-electron system  
using a model potential which provides the correct ground-state ionization
potential for a fixed internuclear distance and behaves like the pure Coulomb
potential of a hydrogen atom for large $r$. The   
total wave function is expanded in a one-center approach in eigenfunctions of 
the one-electron model Hamiltonian of the target. The radial part of the  
basis functions is expanded in B-spline functions and the angular part in 
a symmetry-adapted sum of spherical harmonics. The collision process is 
described with the help of the classical trajectory approximation.  
 
It was found that the ionization cross sections depend approximately linear on
$R_n$ in the interval $1.0\au \le R_n \le 2.11$\au\ \ 
The dependence of $\sigma_{\rm ion}(R_n)$ on $R_n$ diminishes with higher
energies. Cross sections which account
for the vibrational motion of the H$_2$ nuclei can be obtained by employing
closure, exploiting the linear behavior of $\sigma_{\rm ion}(R_n)$, and
performing the calculations at $R_n=\mean{R_n}=1.448$\au\  

The results of the calculations for electron loss in $p$ + H$_2$ collisions
agree  with experimental and theoretical data indicating the applicability of
the used method. 
The present ionization cross sections for \pb + H$_2$ collisions agree for
$E>90$\,keV with the experiment. For $20\,{\rm keV} < E < 80$\,keV 
the magnitude of the calculated $\sigma_{\rm ion}$ is still comparable to the
experimental data, though both curves start to have a different
slope. The calculated excitation cross sections for \pb + H$_2$ collisions were
found to be very similar to those for the excitation of a single hydrogen atom
by antiproton impact. 

An electron-energy spectrum surface $S(\epsilon,E)$ for \pb + H$_2$ collisions
is presented for a wide electron-energy range $0 < \epsilon < 12$\au\ and for
impact energies $48\le E \le 1015$\,keV.  In the interval $\epsilon < 3$\au\
the electron-spectrum curves $S(\epsilon)$ for fixed impact energies $E$ are
smooth 
curves which do not cross. The $S(\epsilon)$ curves are ordered according to the
corresponding impact energy with decreasing magnitude for 
increasing $E$. For higher $\epsilon$ crossings of the $S(\epsilon)$ occur.
Thereby, it is always the uppermost curve which crosses all lower lying
spectrum curves $S(\epsilon)$ which belong to larger $E$. The present
electron-energy spectrum surface $S(\epsilon,E)$ for $p$ + H$_2$ collisions
also includes the electron-capture by the projectile which 
manifests itself in local maxima of the spectrum curves $S(\epsilon)$ for a
given impact energy. The position of the peaks of $S(\epsilon)$ is given by
$\epsilon = E / M_p = (v_p)^2 / 2$. 

A comparison of the used model potential $V_{\rm mod}$ with a scaled hydrogen
atom with comparable ionization potential yields similar ionization cross
sections. Therefore, the ionization process appears to be mainly depending on
the ionization potential. The cross sections for excitation, however, differ
notably which may be explained with the differing binding energies of the
dipole-allowed bound states in both models. Since the excitation energies of
the lowest $p$ states of the model $V_{\rm mod}$ coincide with the
statistically-weighted dipole-allowed excitation energies for the H$_2$
molecule the model $V_{\rm mod}$ is considered to be superior to the 
description with a scaled hydrogen atom.

Concerning the applicability of the used model potential $V_{\rm mod}$ it was
demonstrated that it is suitable for describing 
ionization in \pb + H$_2$ collisions at impact energies $E >
90$\,keV. Furthermore, the model is capable of determining the dependence of
the cross sections on the internuclear distance. Even the calculation of
excitation cross sections seems to be meaningful. Thereby, it has to be
emphasized that besides the one parameter $\alpha$ which is directly determined
by the ionization potential no additional parameter is included in the
potential in order to fit the energies or wave functions to those of the
correct electronic states. On the other hand, not all 
effects which may be of increasing importance at low impact energies can be
described by the model. First, the influence of a second electron is solely
incorporated as a 
screening, second, no dependence on the molecular orientation during the
collision is allowed for and third, vibrational excitation which also includes
dissociation is not considered. Therefore, it would be eligible to perform full
calculations which take the molecular properties of the 
target as well as the two-electron effects, like double ionization or
ionization excitation, into account. Such a theoretical effort, which is
currently in preparation, accompanied with precise measurements at low
antiproton energies would lead to a better understanding of the \pb + H$_2$
collision process for $E<100$\,keV.  

%
%
%

\begin{center} 
 {\bf ACKNOWLEDGMENTS}  
\end{center} 
The authors wish to thank the referee for kindly drawing their attention
to the work of Elizaga {\it et al.}~\cite{sct:eliz99}.
The authors are grateful to BMBF (FLAIR Horizon) and {\it Stifterverband f\"ur 
  die deutsche Wissenschaft} for financial support.

 

\end{document}